\documentclass[a4paper,onecolumn,11pt,accepted=2026-07-31]{quantumarticle}
\pdfoutput=1
\usepackage[utf8]{inputenc}
\usepackage[english]{babel}
\usepackage[T1]{fontenc}
\usepackage{amsmath}
\usepackage{hyperref}
\usepackage{amssymb}
\usepackage{tikz}
\usepackage{lipsum}
\usepackage{mathrsfs}

\begin{document}

\title{Entanglement Breaking Structure of Cartan-Covariant Quantum Channels}

\author{Sean Prudhoe}
\affiliation{Department of Health Science, South College, 3904 Lonas Dr, Knoxville, TN, USA}
\orcid{0009-0009-9586-8817}
\email{sprudhoe@south.edu}

\maketitle

\begin{abstract}
 Cartan-covariant quantum channels are introduced and studied using the Choi-Jamio{\l}kowski isomorphism. The channels are Cartan-covariant as the covariance groups considered form symmetric pairs with the special unitary group SU($n$), and their associated Cartan involution  provides a route to exactly compute the eigenspectrum of their Choi states and the partial transpose for any $n\in \mathbb{N}$. These channels include previously studied SO$(n)$-covariant channels, and further include Sp$(\frac{n}{2})$-covariant channels (when $n$ is even) and S(U($p$) $\times$ U($q$))-covariant channels (where $n\!=\!p\!+\!q$). We show that all Cartan-covariant channels satisfy the PPT$^{2}$-conjecture and further demonstrate the nontrivial nature of this result for the class of Sp$(\frac{n}{2})$-covariant and ${\rm S}({\rm U}(p) \!\times\!{\rm U}(q))$-covariant channels.
 \end{abstract}

\section{\label{sec:intro} Introduction}
Quantum channels play a central role in quantum information theory, describing both the unitary and non-unitary evolution of physical quantum systems \cite{nielsen00,BRE02}. Understanding the structure of quantum channels and whether they are entanglement breaking is essential to design and implement quantum protocols that leverage entanglement as a resource \cite{Horodecki_2003}. However, determining the entanglement breaking structure of generic channels is NP hard because it is equivalent to determining the separability of the Choi state.

An active area of research regarding entanglement breaking channels is the ${\rm PPT}^{2}$-conjecture, proposed by Christandl \cite{Christandl}. The conjecture states that the composition of any completely co-positive quantum channels results in an entanglement-breaking channel. It implies that PPT-entangled states cannot be used as a resource in quantum repeaters or other similar quantum network structures that leverage any non-zero amount of entanglement \cite{Christandl_2017}. Although still an open problem in quantum computation, work in the past decade has proven the PPT$^{2}$-conjecture for limited classes of channels. For example, the PPT$^{2}$-conjecture was proven for low Kraus rank channels, approximately depolarizing channels, and Gaussian channels \cite{Christandl_2019}.

The PPT$^{2}$-conjecture is especially amenable to study over the class of covariant quantum channels. Covariant channels are quantum channels where a group action exists which commutes with the action of the channels \cite{Scutaru_1979}. Explicitly, a unitary representation of a group ($\mathcal{G}$) exists such that  
\begin{equation}
    U_{g}\mathscr{E}(\cdot)U^{\dag}_{g}=\mathscr{E}\left(U_{g}(\cdot)U^{\dag}_{g}\right)
\end{equation}
for all $g\in \mathcal{G}$  where $\mathscr{E}$ is the $\mathcal{G}$-covariant channel. Group symmetry and the study of entanglement breaking of quantum channels pair extremely well. For compact covariance Lie groups, the Haar measure exists and allows for the construction of the twirl operation, 
\begin{align}
    \Xi_{\rm twirl}(\cdot)=\int 
    d\mu_{\rm Harr}U_{g}\otimes V_{g}(\cdot)U^{\dag}_{g}\otimes V^{\dag}_{g}\, 
\end{align}
which projects bipartite matrices onto the space of  $\mathcal{G}$-invariant bipartite matrices. The real power in connection to entanglement breaking is that twirl operation preserves separability \cite{Vollbrecht_2001}. In fact, all entanglement witnesses map to $\mathcal{G}$-invariant entanglement witnesses under the twirl operation.

Recently, it has been shown that quantum channels covariant with respect to the diagonal unitary group satisfy the PPT$^{2}$-conjecture \cite{Singh_2022} and further that the conjecture holds for large, random, diagonal orthogonal covariant channels in \cite{nechita_2024}. A sample of recent work studying related aspects of covariant quantum channels includes \cite{Mozrzymas_2017,Gschwendtner_2021,chru_2024}. To extend on previous work, in this article we consider covariances Lie group such that their Lie algebras ($\mathcal{A}$) form a symmetric pair with the special unitary algebra $\mathfrak{su}(n)=\mathcal{A}\oplus \left(\mathfrak{su}(n)/\mathcal{A}\right)$. The quotient space $\mathcal{B}=\mathfrak{su}(n)/\mathcal{A}$ is known as a symmetric or homogeneous space. By showing that all Sp($\frac{n}{2}$) covariant channels and SO(4) covariant channels satisfy the PPT$^{2}$-conjecture, it follows that all  Cartan-covariant channels satisfy the PPT$^{2}$ conjecture from the results of $\cite{Vollbrecht_2001,Singh_2022}$.

These results, most importantly, on the symplectic covariant channels, are different from existing work for several reason. Chiefly, symplectic covariant channels are never diagonal unitary covariant. The Cartan subalgebra of the unitary group, that generates the diagonal unitary group, is too large be contained within Sp($\frac{n}{2}$). For example assume $n=2m$, then there are $2m-1$ generators in the diagonal unitary group but only $m$ generators in the Cartan subalgebra of the compact symplectic group. Therefore the results of \cite{Singh_2022} do not apply.  However these results are applicable to the S-covariant channels, as S-type unitary covariance implies diagonal unitary covariance. Another useful difference to other work is that we are working with reducible representations of the symmetry groups. For example, useful entanglement breaking conditions were found for channels  covariant with respect to an irreducible representation of a symmetry group $\cite{park2023universalframeworkentanglementdetection}$.  
 \subsection{\label{sec:level2}Quantum channels and the Choi-Jamio\l{}kowski state} 
We begin with a brief review of quantum channels. 
A quantum channel ($\mathcal{E}$) is a map 
\begin{equation}
    \mathscr{E}: \mathbb{M}_{n_{1}}(\mathbb{C})\rightarrow \mathbb{M}_{n_{2}}(\mathbb{C})
\end{equation}
that is trace-preserving and completely-positive, where $\mathbb{M}_{n}(\mathbb{C})$ is the space of $n\times n$ complex matrices. For our purposes, we assume that the input and output spaces are isomorphic $n_{1}=n_{2}=n$. One method of showing a map is completely positive is to construct a Kraus decomposition,
\begin{equation}
\mathscr{E}\left[\rho\right]=\sum_{i}K_{i}\rho K_{i}^{\dag}\,.
\end{equation}
The completely positive map is then trace-preserving if, 
\begin{equation}
\sum_{i}K^{\dag}_{i}K_{i}=\mathbbm{1} \,.
\end{equation}
The Kraus decomposition is not unique as typically many choices of Kraus operators ($K_{i}$) exist. For example, any unitary transformation performed on the set $\{K_{i}\}$ generates a new set of Kraus operators. However, there is a minimal number of Kraus operators necessary to generate a given channel, called the Kraus rank.

The lack of a unique set of the Kraus operators makes the Kraus decomposition unfavorable for parameterizing quantum channels. A better alternative is the use of Choi's theorem and the Choi-Jamio\l{}kowski isomorphism to parameterize a set of quantum channels (for example, \cite{ruskai2001analysis}).  Choi's theorem states that a map $\mathscr{E}$ is completely positive iff the Choi state \cite{Choi_1975}
\begin{equation}
\varrho_{\mathscr{E}}=(\mathbbm{1}\otimes\mathscr{E})\!\left[\omega\right]
\end{equation}
is positive, where $\omega=|\Psi\rangle\!\langle \Psi|$ is the maximally entangled pure state,
\begin{align}
\label{eq:maxent}
    |\Psi\rangle =\sum_{\ell=1}^{n}|\ell\ell\rangle \,. 
\end{align}
The rank of the Choi state is equal to the Kraus rank of the channel, and its eigenvectors with non-zero eigenvalues form a Kraus basis for the corresponding channel.

\subsection{Completely co-positive maps and entanglement breaking}
An additional class of maps, useful in the study of channels, are the completely co-positive maps and the closely related completely co-positive channels. Denoting the transpose operator (in the computational basis) as T$\left(|\ell\rangle\!\langle m|\right)\!=\!|m\rangle \!\langle \ell|$, a map $\mathscr{E}$ is called completely co-positive map if T$\circ\mathscr{E}$ is completely positive. A channel that is completely co-positive is known as a positive partial transpose (PPT) channel, since the Choi state of such a channel has a positive partial transpose, i.e. the partial transpose  
\begin{align}
\widetilde{\varrho}_{\mathscr{E}}=(\mathbbm{1}\otimes{\rm T})\varrho_{\mathscr{E}}
\end{align}
is also a density matrix. The (partial) transpose is important in quantum information, as all separable mixed states have a positive partial transpose. However, the converse is famously known to hold only generically for 2$\times$2 and 2 $\times$3 mixed states \cite{Horodecki_1996}. There are of course subclasses of mixed states in larger dimensions where the PPT states are always separable, perhaps the best known are the Werner states and isotropic states \cite{Werner_1989}.

With respect to channels, a Choi matrix is separable iff its corresponding channel is entanglement breaking. Such channels certainly destroy any entanglement between the quantum system and an ancillary system, as they completely destroy maximal (bipartite) entanglement. A characterization of the parameters of a family of channels where they become entanglement breaking helps determine the feasibility of various quantum protocols and algorithms in a given physical setting. For example, quantifying what region of noise in a quantum circuit is entanglement breaking, allows for the study of surprisingly robust entangled states \cite{Zhang_2013}.   

\section{The Cartan decomposition and derived structures}
\label{sec:SymCov}
 Central to our construction is the notion of a Cartan decomposition of a Lie algebra. Consider the Killing-Cartan metric over $\mathfrak{su}(n)$ defined such that for any two observables, we have 
\begin{align}
\langle\mathcal{O}_{1},\mathcal{O}_{2}\rangle \equiv{\rm tr}\left(\mathcal{O}^{\dag}_{1}\mathcal{O}_{2}\right)\,.
\end{align}
By definition, the Killing-Cartan metic is ad($\mathfrak{su}(n)$)-invariant. Let $G_{\rm i}$ be a basis of observables such that $\langle G_{\rm i},G_{\rm j}\rangle=\delta_{\rm ij}$, with the inclusion of an element $G_{0}\propto\mathbbm{1}$. For the remainder of the article, non-cursive Latin indices in the middle of the alphabet, e.g. i and j, take values from 0 to $n^{2}-1$. The remaining observables $G_{\rm i}$ with i$\neq 0$ are trace-less. A Cartan decomposition is equivalent to the existence of an involution such that 
\begin{equation}
\theta\left([\mathcal{O}_{1},\mathcal{O}_{2}]\right)= [\theta\!\left(\mathcal{O}_{1}\right),\theta\!\left(\mathcal{O}_{2}\right)]\,,
\end{equation}
for any observables $\mathcal{O}_{1}$,$\mathcal{O}_{2}\in\mathfrak{su}(n)$.
The map $\theta$ is a Lie algebra automorphism known as a Cartan involution. Let ($\{A_{r}\}$,$\{B_{\nu}\}$) be an orthonormal basis of the Killing-Cartan metric such that for all $r$ and $\mu$ we have
\begin{equation}
\begin{split}
    \theta\!\left(A_{r}\right)&=A_{r} \\
    \theta\!\left(B_{\mu}\right)&=-B_{\mu}
\end{split}
\end{equation}
i.e. an orthonormal basis that consists of eigenvectors of $\theta$. Latin indices in the middle of the alphabet (e.g. $r$) are in the set $\{1,2,...,a\}$, while Greek indices (e.g. $\mu$) are in the set $\{a+1,a+2,...,n^{2}-1\}$, where $a$ is the dimension of $\mathcal{G}$ as a topological manifold. Denoting $\mathcal{A}={\rm span}\left(A_{r}\right)$ and $\mathcal{B}={\rm span}\left(B_{\mu}\right)$, the following commutation relations are satisfied 
\begin{align}
\label{eq:Cartan}
&[\mathcal{A},\mathcal{A}]\subset \mathcal{A} \nonumber \\
&[\mathcal{A},\mathcal{B}]\subset \mathcal{B} \nonumber \\
&[\mathcal{B},\mathcal{B}]\subset \mathcal{A} \,. 
\end{align}
Therefore, $\mathcal{A}$ is a Lie subalgebra of $\mathfrak{su}(n)$, and $\mathcal{B}$ forms a representation of $\mathcal{A}$. For the Cartan decompositions considered in this work, the stronger statement that $\mathcal{B}$ is an irreducible (real) representation of $\mathcal{A}$ holds (\cite{dalessandro2005quantumsymmetriescartandecompositions}). 

Without assuming any further properties of $\mathcal{A}$, the representations $(\mathbbm{1},\mathcal{A},\mathcal{B})$ are the only  $\mathcal{A}$-invariant subspaces within $\mathfrak{su}(n)$. 
However, when $\mathcal{G}$ is not simple, $\mathcal{A}$ splits further into smaller invariant subspaces 
\begin{align}
\mathcal{A}&=\bigoplus_{I}\mathcal{A}_{I} 
\end{align}
where $[\mathcal{A}_{I},\mathcal{A}_{J}]=0$ when $I\neq J$. 
The subalgebra $\mathcal{A}$ is not simple for the covariance groups $\mathcal{G}$= S(U$(p)\times$ U$(q)$) and $\mathcal{G}$= SO(4), where $\mathcal{A}=\mathfrak{su}(p)\oplus\mathfrak{su}(q)\oplus\mathfrak{u}(1)$ and $\mathcal{A}=\mathfrak{so}(3)\oplus \mathfrak{so}(3)$, respectively. What remains to fully parameterize a Cartan-covariant channel is the center ($\mathcal{G}_{\mathcal{C}}$) of the covariance group. A unitary $U\in {\rm U}(n)$ is an element of $\mathcal{G}_{\mathcal{C}}$ iff  
\begin{align}
U\mathcal{G}_{a}U^{\dag}=\mathcal{G}_{a}
\end{align}
for all $\mathcal{G}_{a}\in \mathcal{G}$. $\mathcal{G}_{\mathcal{C}}$ is a closed Lie subgroup of $\mathcal{G}$ and is generated by $\mathcal{C}\subset\mathcal{A}$ \cite{Kobayashi_1963}.  Obviously, $U=\mathbbm{1}$ is always an element of the center, and $\mathcal{C}=0$ when $\mathbbm{1}$ is the only central element. In such a case, the center is called trivial. SO($n$) and Sp($\frac{n}{2}$) both have trivial centers for all admissible $n$, while the group S(U($p$)$\times$U($q$)) contains a single non-trivial element for any choice of $n>p\geq q>0$. 
\subsection{Cartan-covariant quantum channels}
\label{sec:CCQC}
Assume that the special unitary group has the following orthogonal decomposition 
\begin{align}
    \mathfrak{su}(n)=\mathcal{A}\oplus\mathcal{B}=\left(\bigoplus_{I}\mathcal{A}_{I}\right) \oplus \mathcal{B} \oplus \mathcal{C}
\end{align}
where $\left\{\mathcal{A},\mathfrak{su}(n)\right\}$ forms a symmetric pair and where the central generators are explicitly isolated from $\mathcal{A}$. Each of the $\mathcal{A}$-invariant subspaces is spanned by the generators 
\begin{align}
\label{eq:CartanBasis}
    \mathcal{A}_{I}&={\rm span}\left(A_{r_{I}}\right) \nonumber \\
    \mathcal{B}&={\rm span}\left(B_{\mu}\right) \nonumber \\
    \mathcal{C}&={\rm span}\left(C_{M}\right) \,,
\end{align} 
which are eigenvectors of the Cartan involution that form an orthonormal basis of the Killing-Cartan metric. It is important to note that each $C_{M}$ generates a trivial representation of $\mathcal{G}$, where $M$ indexes over the generators of the non-trivial central elements. The set of generators, listed in the order of Eq.(\ref{eq:CartanBasis}), with the inclusion of the identity operator, is referred to as a Cartan basis in the following. A Cartan-covariant channel $\mathscr{E}_{\mathcal{A}}$, which is covariant with respect to the Lie group $\mathcal{G}=\exp{i\mathcal{A}}$, has the following representation in the Cartan basis,  
\begin{align}
\mathscr{E}_{\mathcal{A}}=\begin{pmatrix}
        1&\vec{0} \\
\vec{\mathscr{S}}_{\mathcal{A}}&\mathscr{R}_{\mathcal{A}}\mathscr{D}_{\mathcal{A}}
    \end{pmatrix}
\end{align}
where $\mathscr{R}_{\mathcal{A}}$ is a $(n^{2}-1)\times(n^{2}\!-\!1)$ rotation matrix, $\mathscr{D}_{\mathcal{A}}$ is a $(n^{2}\text{-}1)\times(n^{2}\text{-}1)$ diagonal matrix, and $\vec{\mathscr{S}}_{\mathcal{A}}$ is a column vector of dimension $(n^{2}\text{-}1)$. For generic quantum channels, the matrix $\mathscr{D}_{\mathcal{A}}$ is symmetric but not necessarily diagonal. However, due to Cartan-covariance, whatever rotational freedom allows for a more general $\mathscr{D}_{\mathcal{A}}$ may simply be absorbed into the definition of $\mathscr{R}_{\mathcal{A}}$ without loss of generality.

In order for $\mathscr{E}_{\mathcal{A}}$ to be Cartan-covariant, it must commute with all elements of the covariance group. The covariance transformations in the Cartan basis are rotation matrices, $\mathscr{R}_{\mathcal{G}}$, and $\mathcal{G}$-covariance is enforced through 
\begin{align}
\mathscr{R}_{\mathcal{G}}\mathscr{R}_{\mathcal{A}}&=\mathscr{R}_{\mathcal{A}}\mathscr{R}_{\mathcal{G}} \nonumber \\
\mathscr{R}_{\mathcal{G}}\mathscr{D}_{\mathcal{A}}&=\mathscr{D}_{\mathcal{A}}\mathscr{R}_{\mathcal{G}} \nonumber \\
\mathscr{R}_{\mathcal{G}}\vec{\mathscr{S}}_{\mathcal{A}}&=\vec{\mathscr{S}}_{\mathcal{A}}
\end{align}
for all $\mathscr{R}_{\mathcal{G}}\in {\rm Ad}\mathcal{G}$. From the first line, we must conclude that $\mathscr{R}_{\mathcal{A}}\in {\rm Ad}\mathcal{G}_{\mathcal{C}}$ and therefore takes the generic form 
\begin{align} 
\mathscr{R}_{\mathcal{A}}(\cdot)&=\exp{\left(-i\sum_{M}\phi_{M}\mathcal{C}_{M}\right)}(\cdot)\exp{\left(i\sum_{M}\phi_{M}\mathcal{C}_{M}\right)}\,, 
\end{align}
while the last line implies that $\vec{\mathscr{S}}_{\mathcal{A}}$ lies entirely within $\mathcal{C}$ that is,  
\begin{align}
\mathscr{E}_{\mathcal{A}}\left(\mathbbm{1}\right)&=\mathbbm{1}+\sum_{M}\delta_{M}C_{M} \,.
\end{align}
Therefore, if $\mathcal{G}_{\mathcal{C}}$ is trivial, then $\mathscr{E}_{\mathcal{A}}$ is unital and only has a trivial unitary component. And finally, from the second covariance relation, we must conclude that $\mathscr{D}_{\mathcal{A}}$ is proportional to the identity operator over each $\mathcal{A}$-invariant subspace, that is, 
\begin{align}
\mathscr{D}_{\mathcal{A}}=\mathscr{A}_{\mathcal{A}}\oplus\mathscr{B}_{\mathcal{A}}\oplus\mathscr{C}_{\mathcal{A}} 
\end{align}
where $\mathscr{A}_{\mathcal{A}}$, $\mathscr{B}_{\mathcal{A}}$, and $\mathscr{C}_{\mathcal{A}}$ are defined as
\begin{align}
\mathscr{A}_{\mathcal{A}}=\sum_{I}\alpha_{I}\mathbbm{1}_{\mathcal{A}_{I}} ,\quad \mathscr{B}_{\mathcal{A}}=\beta \mathbbm{1}_{\mathcal{B}}, \quad \mathscr{C}_{\mathcal{A}}=\sum_{M}\gamma_{M}\mathbbm{1}_{\mathcal{C}_{M}}\,. 
\end{align}

\subsection{Cartan-invariant Choi matrices}
In the previous section, we parametrized the action of Cartan-covariant channels over observables. However,  parameterizing instead the Cartan-invariant Choi matrices allows for a much simpler computation of the eigenvalues. The connection between covariant channels and invariant states is simple to show, and we sketch the connection briefly here. Suppose that we have a $\mathcal{G}$-covariant channel, then it follows that 
\begin{align*}
    (\mathbbm{1}\otimes V_{\mathcal{G}})(\mathbbm{1}\otimes\mathscr{E})\!\left[\omega\right] (\mathbbm{1}\otimes V^{\dag}_{\mathcal{G}})&=(\mathbbm{1}\otimes\mathscr{E})\left[(\mathbbm{1}\otimes V_{\mathcal{G}})\omega(\mathbbm{1}\otimes V^{\dag}_{\mathcal{G}})\right] \\ &=(\mathbbm{1}\otimes\mathscr{E})\left[( V^{\rm T}_{\mathcal{G}}\otimes\mathbbm{1})\omega( \bar{V}_{\mathcal{G}}\otimes\mathbbm{1})\right] \\
    &=( V^{\rm T}_{\mathcal{G}}\otimes\mathbbm{1})(\mathbbm{1}\otimes\mathscr{E})\!\left[\omega\right]( \bar{V}_{\mathcal{G}}\otimes\mathbbm{1}) \,,
\end{align*}
where $\bar{V}$ denotes that the complex conjugate in the basis $|\Psi\rangle$ was expressed in Eq.(\ref{eq:maxent}).
The second line follows from  the ricochet property of the maximally entangled state \cite{biamonte2020lecturesquantumtensornetworks}, and the last line follows from the $\mathcal{G}$-covariance of the identity channel. Multiplying both sides by the appropriate inverses and setting $U_{\mathcal{G}}=\bar{V}_{\mathcal{G}}$, we find that 
\begin{equation}
    \varrho_{\mathscr{E}}=(U_{\mathcal{G}}\otimes \bar{U}_{\mathcal{G}})\varrho_{\mathscr{E}}(U^{\dag}_{\mathcal{G}}\otimes U^{\rm T}_{\mathcal{G}})\,.
\end{equation}
 Therefore, the Choi state of a $\mathcal{G}$-covariant channel is  Ad$\left(\mathcal{G}\otimes \bar{\mathcal{G}}\right)$-invariant. Following the standard convention, e.g. \cite{Vollbrecht_2001,chru_2024}, we refer to these states as $\mathcal{G}$-invariant (bipartite) states. For any covariance group $\mathcal{G}$ the maximally mixed and entangled states, 
\begin{align}
\mathbbm{1}&=\sum_{\ell m}|\ell m\rangle\!\langle \ell m| \nonumber \\
\omega&=\sum_{\ell m}|\ell \ell \rangle\!\langle m m|
 \end{align}
are always $\mathcal{G}$-invariant (unnormalized) states. By further assuming Cartan-invariance, there is also the additional structure 
\begin{equation}
\widetilde{\omega}=\left(\mathbbm{1}\otimes\theta\!\left[\omega\right]\right)\,.
\end{equation}
 These invariant components arise from the action of the depolarizing, identity, and Cartan-involution maps, respectively. When $\mathcal{G}$ is simple and has a trivial center, these structures exhaust the possible invariants. However, for covariance groups that are not simple and/or do not have a trivial center, include additional invariant structures. We investigate these cases in detail later in this article.  
\subsubsection{Ex: SO($n$)-invariant Choi matrices}
To illustrate the techniques used in this article, we consider SO($n$)-invariant Choi matrices previously studied in \cite{Vollbrecht_2001,Chru_ci_ski_2006}. The Cartan involution is $\theta(\cdot)=-(\cdot)^{\rm T}$ with, 
\begin{align}
\widetilde{\omega}=-\sum_{\ell m}|\ell m\rangle\!\langle m \ell | \,, 
\end{align}
i.e. the Cartan-involution essentially generates the flip operator under the Choi isomorphism. A generic SO($n$)-invariant Choi matrix ($n\neq$ 4) is given by, 
\begin{align}
\varrho_{\mathcal{A}}=\Gamma\mathbbm{1}+\Omega\omega+\widetilde{\Omega}\widetilde{\omega}\,,
\end{align}
where $n\Gamma=1-\left(\Omega-\widetilde{\Omega}\right)$ so that tr$\varrho_{\mathcal{A}}=n$, ensuring that the channel is trace-preserving. The partial transpose is, 
\begin{align}
\widetilde{\varrho}_{\mathcal{A}}=-(\mathbbm{1}\otimes\theta)\left[\varrho_{\mathcal{A}}\right]=\Gamma\mathbbm{1}-\widetilde{\Omega}\omega-\Omega\widetilde{\omega}\,.
\end{align}
Part of the simplicity of orthogonal-channels is that $[\varrho_{\mathcal{A}},\widetilde{\varrho}_{\mathcal{A}}]=0$, therefore, once the eigenvalues for the Choi state are computed, the eigenvalues of the partial transpose are obtained through a simple change of parameters. 
The diagonalization of the Choi matrix is done by constructing projectors onto each SO$(n)$-invariant subspace. Using the algebra of SO($n$)-invariants,
\begin{align}
    \omega^{2}&=n\omega \nonumber \\
\widetilde{\omega}^{2}&=\mathbbm{1} \nonumber \\
\omega\widetilde{\omega}&=\widetilde{\omega}\omega=-\omega\,, 
\end{align}
the following SO$(n)$-invariant projectors are constructed
\begin{align}
\Pi_{\mathbbm{1}}&=\frac{\omega}{n} \nonumber \\  
\Pi_{\mathcal{A}}&=\frac{1}{2}\left(\mathbbm{1}+\tilde{\omega}\right)\nonumber \\ 
\Pi_{\mathcal{B}}&= \frac{1}{2}\left(\mathbbm{1}-\widetilde{\omega}\right)-\frac{\omega}{n} \,. 
\end{align}
Defining $n{\rm w}={\rm tr}\left(\omega \varrho_{\mathcal{A}}\right)$ and $n\widetilde{\rm w}=-{\rm tr}\left(\widetilde{\omega} \varrho_{\mathcal{A}}\right)$ and using tr$\left(\varrho_{\mathcal{A}}\Pi_{\lambda}\right)={\rm tr}\left(\Pi_{\lambda}\right) \!\lambda$, we find the completely positive constraints for orthogonal channels to be,  
\begin{align}
\lambda_{\mathbbm{1}}& ={\rm w} \geq 0 \nonumber \\
\lambda_{\mathcal{A}}&= \frac{1}{2}\left(1-\widetilde{\rm w}\right) \geq0\nonumber \\
\lambda_{\mathcal{B}}&=\frac{1}{2}\left(1+\widetilde{\rm w}\right)-\frac{\rm w}{n}\geq0 \,, 
\end{align}
and the completely co-positive constraints are found by the exchange w$ \leftrightarrow \widetilde {\rm w}$. These are the same results as those derived in \cite{Vollbrecht_2001}, where the authors went on to show that $\widetilde{\varrho}_{\mathcal{A}}\geq0$ iff $\varrho_{\mathcal{A}}$ is separable. In terms of channels, this means that $\mathscr{E}_{\mathcal{A}}$ is entanglement breaking whenever it is completely co-positive. Therefore, for orthogonal-covariant channels, the PPT$^{2}$-conjecture holds trivially, as the stronger statement EB=PPT holds. In Appendix B, we explore the exceptional case of SO(4), which is the only non-simple special orthogonal group. This adds an additional parameter to the space of covariant channels and we study the entanglement breaking properties. We explicitly show that EP=PPT hold over this larger class of covariant channels.
\section{Sp($\frac{n}{2})$-invariant Choi matrices}
\begin{figure}
    \centering
\includegraphics[width=0.5\linewidth]{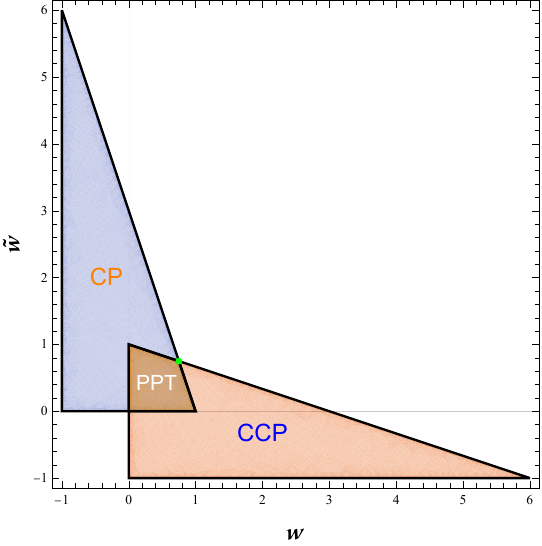}
    \caption{The regions of Sp(3)-covariant completely positive (blue) and completely co-positive (orange) maps are shown. The intersection of the CP and CCP regions, the PPT channels, is a kite, and the only $n$ dependence of appears in the vertex labeled by green mark. The location of this extreme PPT channel for a generic $n$ is ($\frac{n}{n+2}$,$\frac{n}{n+2}$).}
\label{fig:symPPTreg}
\end{figure}

In the last section, we derived previously known results about SO($n$)-covariant channels using the notion of Cartan-invariance. We now prove new results for the set of Sp$(\frac{n}{2})$-covariant quantum channels. The Cartan involution is $\theta(\cdot)=-\mathcal{J}(\cdot)^{\rm T}\mathcal{J}^{\dag}$ where $\mathcal{J}$ is the symplectic form defined as
\begin{align}
    \mathcal{J}=\begin{pmatrix}
        0 & \mathbbm{1}_{\frac{n}{2}} \\
        -\mathbbm{1}_{\frac{n}{2}} & 0 
    \end{pmatrix} \,. 
\end{align}
As the compact symplectic group is simple and has a trivial center, the most generic form of Sp($\frac{n}{2}$)-invariant trace-preserving Choi matrix is, 
\begin{align}
\label{eq:Choisym}
\varrho_{\mathcal{A}}=\Gamma\mathbbm{1}+\Omega\omega+\widetilde{\Omega}\widetilde{\omega} \,,
\end{align}
where $\widetilde{\omega}$ is defined using the Cartan-involution associated with the symplectic group and $n\Gamma=1-\left(\Omega-\widetilde{\Omega}\right)$. Because the Cartan involution involves the transpose, the partial transpose of the Choi state can be computed as 
\begin{align}
\widetilde{\varrho}_{\mathcal{A}}=-(\mathbbm{1}\otimes\theta)\left[\varrho_{\mathcal{A}}\right]=\Gamma\mathbbm{1}-\widetilde{\Omega}\omega-\Omega\widetilde{\omega} \,. 
\end{align}
The algebra of Sp$(\frac{n}{2})$-invariant structures is 
\begin{align}
    \omega^{2}&=n\omega \nonumber \\
\widetilde{\omega}^{2}&=\mathbbm{1} \nonumber \\
\omega\widetilde{\omega}&=\widetilde{\omega}\omega=\omega\,,
\end{align}
where the first difference with the SO($n$)-invariant case emerges; the sign of $\omega$ in the third line is positive. We see from Fig.(\ref{fig:symPPTreg}) that this causes a dependence on $n$  to appear in the shape of the region of completely positive and completely co-positive channels (PPT region). Using the above algebra of invariants, the following complete set of Sp$(\frac{n}{2})$-invariant projectors is constructed, 
\begin{align}
\Pi_{\mathbbm{1}}&=\frac{\omega}{n} \nonumber \\  
\Pi_{\mathcal{A}}&=\frac{1}{2}\left(\mathbbm{1}-\widetilde{\omega}\right)\nonumber \\ 
\Pi_{\mathcal{B}}&= \frac{1}{2}\left(\mathbbm{1}+\widetilde{\omega}\right)-\frac{\omega}{n} \,. 
\end{align}
The completely positive constraints are then computed from the overlaps of these projectors with the Choi matrix. We find,  
\begin{align}
\lambda_{\mathbbm{1}}&={\rm w} \geq 0 \nonumber \\
\lambda_{\mathcal{A}}&= \frac{1}{2}\left(1+\widetilde{\rm w}\right) \geq0\nonumber \\
\lambda_{\mathcal{B}}&=\frac{1}{2}\left(1-\widetilde{\rm w}\right)-\frac{\rm w}{n}\geq0 \,, 
\end{align}
where, again, the completely co-positive constraints are found by exchanging w and $\widetilde{\rm w}$ and w and $\widetilde{\rm  w}$ are have the same definition as in the  preceding section. 
 From the completely positive and completely co-positive constraints, the region of PPT symplectic channels is determined and plotted in Fig.(\ref{fig:symPPTreg}). 

 Notice that the compact symplectic group is also simple. Therefore the twirling operation used in \cite{Vollbrecht_2001} exists and has the same behavior as the case of $\mathcal{A}=\mathfrak{so}(n)$. It follows that their techniques can be applied to the problem of determining the separability of symplectic invariant matrices. That is we, we must find a collection of pure states ($\{|\psi_{1},\psi_{2}\rangle\}$) that reproduce the expectation values of $\omega$ and $\widetilde{\omega}$ defining the extreme points of the PPT region. The expectation values w and $\widetilde{\rm w}$, assuming pure inputs, are computed as  
\begin{align}
\label{eq"EBCSP}
{\rm w}&=|\langle\psi_{1}|\bar{\psi}_{2}\rangle|^{2} \nonumber \\
   \widetilde{\rm w}&= |\langle\psi_{1}|\mathcal{J}|\psi_{2}\rangle|^{2} \,,
    \end{align}
where $|\bar{\psi}\rangle$ is the complex conjugate in the computational basis.
In addition to the computational basis states, it is useful to introduce the following set of pure states,
\begin{align}
|x^{\pm}_{\ell m}\rangle&=\frac{1}{\sqrt{2}}\left(|\ell\rangle\pm|m\rangle\right) \nonumber \\
|y^{\pm}_{\ell m}\rangle&=\frac{1}{\sqrt{2}}\left(|\ell\rangle\pm i|m\rangle\right) \,.
\end{align}
These pure states have the following useful properties. The $x$-labeled states transform under index exchange as $|x^{\pm}_{\ell m}\rangle =\pm|x^{\pm}_{m \ell }\rangle$. The $y$-labeled states  transform under index exchange as $|y^{\pm}_{\ell m}\rangle=\pm i|y^{\mp}_{m \ell}\rangle$. For $\ell+m\neq n+1$, $\mathcal{J}$ sends both the $x$-labeled and $y$-labeled states to orthogonal states with the same label, but potentially with a different upper sign. For $\ell+m=n+1$, $\mathcal{J}$ flips the upper sign of the $x$-labeled states and fixes (up to phase) the $y$-labeled states. And finally, the $x$-labeled states remain unchanged by complex conjugation, while the $y$-labeled states change upper sign. In what follows, we construct sets of linearly independent pure states that reproduce the values of $w$ and $\widetilde{w}$ for the extreme PPT channels. By considering the rank of these channels, we know the minimum amount of such pure states required to completely reproduce these channels. 
 
 We start with the Choi matrix corresponding to the vertex $(1,0)$, and we see that the Choi matrix is full rank as it has no non-zero eigenvalues. Therefore, we need to be able to find $n^{2}$ linearly independent pure states, all having $w=1$ and $\widetilde{w}=0$. The computational basis states $|\ell \ell \rangle$ and the $x$-labeled states $|x^{\pm}_{\ell m},x^{\pm}_{\ell m}\rangle$ are easily seen to have the correct expectation values. A quick computation shows that there are $n+n(n-1)=n^{2}$ of these pure states, and further verification shows that they are linearly independent. At the opposite vertex $(0,1)$ the Choi matrix is proportional to $\hat{\Pi}_{\mathcal{A}}$, and therefore rank$\left[\varrho_{\mathcal{A}}\right]=\frac{n(n+1)}{2}$. The states $|\ell \ell'\rangle$ with $\ell'=n-\ell+1$ and $|x^{\pm}_{\ell m},x^{\mp}_{\ell m}\rangle$ have the correct expectation values and are numerous enough to reproduce the Choi state. The extreme channel at the origin has a zero eigenvalue and so has a rank of $n^{2}-1$. The states $|y^{+}_{\ell m},y^{+}_{\ell m}\rangle$ with $\ell\neq m$ and $\ell+m\neq n+1$, and the states $|y^{+}_{\ell \ell'},y^{+}_{m m'}\rangle$ with $\ell\neq m$ and $\ell\neq m'$ have the correct expectation values and are numerous enough to reproduce the Choi state with $w=\widetilde{w}=0$.

All that remains to show is whether the extreme PPT channel marked by the green dot supports pure states, as we have already shown that all other extreme PPT channels do. For this to be true, a collection of pure states must be found such that w$=\widetilde{\rm w}=\frac{n}{n+2}$. The first equality in Eq.(\ref{eq"EBCSP}) implies that $|\psi_{1}\rangle$ needs to have an overlap with $|\bar{\psi}_{2}\rangle$,
\begin{align}
    |\psi_{1}\rangle=\sqrt{\frac{n}{n+2}}|\bar{\psi_{2}}\rangle+\sqrt{\frac{2}{n+2}}|\bar{\psi}^{\perp}_{2}\rangle,
\end{align}
where $|\bar{\psi}^{\perp}_{2}\rangle$ is any normalized state in the orthogonal complement of $|\bar{\psi}_{2}\rangle$. Now, it is straightforward to show that the second equality can never be satisfied when $n>2$. For example, fix $|\psi_{2}\rangle=|\ell\rangle$. To produce the largest possible non-zero matrix element of $\mathcal{J}$, set $|\bar{\psi}^{\perp}_{2}\rangle=|n-\ell+1\rangle$, and find that 
\begin{align}
    |\langle \psi_{1}|\mathcal{J}|\psi_{2}
\rangle| =\sqrt{\frac{2}{n+2}}|\langle n-\ell+1|\mathcal{J}|\ell\rangle|=\sqrt{\frac{2}{n+2}}<\sqrt{\frac{n}{n+2}}\,.
\end{align}
It follows from the linearity of the expectation value that no choice of $|\psi_{2}\rangle$ can be made to satisfy this equality. Therefore, there are no pure states that support the expectation values w$=\widetilde{\rm w}=\frac{n}{n+2}$. Thus, symplectic covariant entanglement breaking channels are a proper subset of the symplectic covariant PPT channels. However, the symplectic channels satisfy the PPT$^{2}$-conjecture, as we now prove. The composition of any two symplectic channels is 
\begin{align}
\varrho_{\mathscr{E}_{2}\mathscr{E}_{1}}&=(\mathbbm{1}\otimes\mathscr{E}_{2})\varrho_{\mathscr{E}_{1}}=\Gamma_{1}(\mathbbm{1}\otimes\mathscr{E}_{2})\left[\mathbbm{1}\right]+\Omega_{1}(\mathbbm{1}\otimes\mathscr{E}_{2})\left[\omega\right]+\widetilde{\Omega}_{1}(\mathbbm{1}\otimes\mathscr{E}_{2})\left[\widetilde{\omega}\right] \nonumber \\ &=\left(\Gamma_{1}+\Gamma_{2}(\Omega_{1}-\widetilde{\Omega}_{1})\right)\mathbbm{1}+\left(\Omega_{1}\Omega_{2}+\widetilde{\Omega}_{1}\widetilde{\Omega}_{2}\right)\omega+\left(\Omega_{1}\widetilde{\Omega}_{2}+\widetilde{\Omega}_{1}\Omega_{2}\right)\widetilde{\omega} \,,
\end{align}
where the unitality of the symplectic channels was used as well as the fact that they commute with $\theta$. It is also useful to determine $\Omega=\Omega({\rm w},\widetilde{\rm w})$ and $\widetilde{\Omega}=\widetilde{\Omega}({\rm w},\widetilde{\rm w})$, where we find 
\begin{align}
    \Omega&=\frac{-1+(n-1){\rm w}+\widetilde{\rm w}}{(n+1)(n-2)} \nonumber \\ 
    \widetilde{\Omega}&=\frac{1-{\rm w}-(n-1)\widetilde{\rm w}}{(n+1)(n-2)} \,.
\end{align}
All that remains to show (more details in Appendix \ref{sec:SymPPT2}) is that the square of the extreme PPT channel (represented by the green dot in Fig.(\ref{fig:symPPTreg}) is contained within the EB region. The channel parameters are $\Omega=-\widetilde{\Omega}=\frac{1}{(n+2)}$, and taking the square, we find w$=\widetilde{\rm w}=\frac{3n+2}{(n+2)^{2}}$. To be in the entanglement breaking region we must have w$=\widetilde{\rm w}\leq\frac{1}{2}$, which holds for all $n\geq2$. Thus, symplectic channels non-trivially satisfy the PPT$^{2}$-conjecture. 
\section{S(U($p$)$\times$U($q$))-invariant Choi matrices}
\label{Section S}
The covariant channels with respect to the S-type unitary group $\mathcal{G}$ = S (U($p$)$\times$U($q$)), are much more complicated than the Cartan-covariant channels seen above. This is because the S-type unitary group is not simple and also has a non-trivial center. The Cartan involution for the S-type unitary group is $\theta(\cdot)=-U(\cdot)U^{\dag}$ where $U=\mathbbm{1}_{p}\oplus(-\mathbbm{1}_{q})$ with $p+q=n$. There is obviously more than one partition $(p,q)$, and for each choice the covariance Lie algebra is $\mathcal{A}=\mathfrak{su}(p)\oplus\mathfrak{su}(q)\oplus\mathfrak{u}(1)$. The first two subalgebras are spanned by the generalized Gell-mann matrices of the form
\begin{align}
    \mathcal{G}_{r} =\begin{pmatrix} \mathcal{G}_{r_{+}}& 0^{\rm T} \\ 0 & \mathcal{G}_{r_{-}}
    \end{pmatrix}
\end{align}
where $\{\mathcal{G}_{r_{+}}\}$ and $\{\mathcal{G}_{{r_{-}}}\}$ are sets of generalized Gell-mann matrices that span $\mathfrak{su}(p)$ and $\mathfrak{su}(q)$ respectively. The last sub-algebra is spanned by the trace-less part of $U$. The remaining generalized Gell-Mann matrices, from $\mathfrak{su}(n)$, span $\mathcal{B}$. 

It is straightforward to construct the set of $\mathcal{G}$-invariant structures. Begin with the maximally entangled state and decompose it into, 
\begin{align}
|\Psi\rangle=\sum_{1}^{p}|\ell_{+}\ell_{+}\rangle+\sum_{p+1}^{n}|\ell_{-}\ell_{-}\rangle=|\Psi_{p}\rangle+|\Psi_{q}\rangle \,. 
\end{align}
Both $|\Psi_{p}\rangle$ and $|\Psi_{q}\rangle$ are obviously invariant under the corresponding S-type unitary group. We use them to define the following S-invariant structures
\begin{align}
\mathcal{P}&= |\Psi_{p}\rangle\!\langle \Psi_{p}| \quad \mathcal{R}=|\Psi_{q}\rangle\!\langle \Psi_{p}| \nonumber\\ 
\mathcal{Q}&=|\Psi_{q}\rangle\!\langle \Psi_{q}| \quad \mathcal{R}^{\dag}=|\Psi_{p}\rangle\!\langle \Psi_{q}|\nonumber \,.
\end{align}
This (incomplete) algebra of invariants is non-abelian, in fact, these four invariants form a Pauli algebra  
\begin{align}
    \mathcal{X}&=\frac{1}{\sqrt{4pq}}(\mathcal{R}+\mathcal{R}^{\dag}) \quad \mathcal{Z}=\frac{1}{2p}\mathcal{P}-\frac{1}{2q}\mathcal{Q} \nonumber \\
    \mathcal{Y}&= \frac{i}{\sqrt{4pq}}(\mathcal{R}-\mathcal{R}^{\dag})\quad   \mathcal{T}=\frac{1}{2p}\mathcal{P}+\frac{1}{2q}\mathcal{Q} \nonumber   \,, 
\end{align}
with normalization $\mathcal{X}^{2}=\mathcal{Y}^{2}=\mathcal{Z}^{2}=\mathcal{T}^{2}=\frac{1}{2}\mathcal{T}$. The remaining invariant structures needed to complete the algebra are $\mathbbm{1}$, $\mathcal{U}_{\rm R}=\mathbbm{1}\otimes U$, $\mathcal{U}_{\rm L}=U\otimes\mathbbm{1}$ and $\mathcal{V}=U\otimes U$. Therefore, the most generic S-invariant Choi matrix is
\begin{align}
\varrho_{\mathcal{A}}={\rm w}\mathbbm{1}+{\rm x}\mathcal{X}+{\rm y}\mathcal{Y}+{\rm z}\mathcal{Z}+{\rm  t}\mathcal{T}+{\rm u_{\rm R}}\, \mathcal{U}_{\rm R}+{\rm u_{\rm L}}\, \mathcal{U}_{\rm L}+{\rm v}\mathcal{V} \,,
\end{align}
and is trace-preserving when tr$_{\rm L}\varrho_{\mathcal{A}}=\mathbbm{1}_{\rm R}$ \cite{watrous2025understandingquantuminformationcomputation}. We find that the S-invariant Choi matrix is trace-preserving, under the additional assumptions 
\begin{align}
   \left(\frac{p+q}{4pq}\right)\!{\rm t}-\left(\frac{p-q}{4pq}\right)\!{\rm z}+(p+q){\rm w}+(p-q){\rm u_{\rm L}}&=1 \nonumber \\
  -\left(\frac{p-q}{4pq}\right)\!{\rm t}+\left(\frac{p+q}{4pq}\right)\!{\rm z}+(p-q){\rm v}+(p+q){\rm u_{\rm R}}&=0 \,.
\end{align}
Therefore, trace-preserving, S-invariant Choi matrices have six free parameters for any $(p,q)$.  Since we are interested in the non-unitary properties of the S-covariant channels, we set y$=\!\!0$. This is equivalent to setting $\phi=0$ in $N=\exp{i \phi U}$. The action of this operator fixes all invariant structures except $\mathcal{X}$ and $\mathcal{Y}$, which it mixes. That leaves 5 free non-unitary parameters, which agrees with the parameter count from our construction of Cartan-covariant channels above. There are 4 depolarizing parameters, as there are four irreducible representations of the group present.  And there is one  shift parameter, since the center has a single non-trivial.
Unlike the previously studied channels, the S-channels are non-unital. The unital conditions are found by enforcing tr$_{\rm R}\varrho_{\mathcal{A}}=\mathbbm{1}_{\rm L}$, and are the trace-preserving conditions with ${\rm u_{\rm L}}$ and ${\rm u_{\rm R}}$ swapped. Checking the independence of this set of four equations, finds that there are only three independent conditions.

The Cartan involution for the S-type unitary group does not include the transpose, so the partial transpose generates an entirely distinct algebra of $\widetilde{\rm S}$-invariants. The partial transpose of the Choi state is as follows,
\begin{align}
\widetilde{\varrho}_{\mathcal{A}}={\rm w}\mathbbm{1}+{\rm x}\widetilde{\mathcal{X}}+{\rm y}\widetilde{\mathcal{Y}}+{\rm z}\widetilde{\mathcal{Z}}+{\rm t}\widetilde{\mathcal{T}}+{\rm u_{\rm R}}\, \mathcal{U}_{\rm R}+{\rm u_{\rm L}}\, \mathcal{U}_{\rm L}+{\rm v}\mathcal{V}
\end{align}
where it no longer holds that$[\varrho_{\mathcal{A}},\widetilde{\varrho}_{\mathcal{A}}]=0$. Therefore, the Choi state and its partial transpose must be diagonalized separately i.e. we must construct S-invariant projectors and $\widetilde{{\rm S}}$-invariant projectors independently. The partial transpose (or equivalent operations) does not permute the $\mathcal{G}$-invariant projectors as was the case for the previously considered Cartan-covariant families. 

Using the algebra of S-invariants (see  Appendix \ref{AppC}), the following set of S-invariant projectors is constructed 
\begin{align}
    \Pi^{\pm}_{\mathbbm{1}}&=\mathcal{T}\pm\left(\cos{\zeta}\mathcal{Z}+\sin{\zeta}\mathcal{X}\right)\nonumber \\ \Pi^{\pm}_{\mathcal{A}}&=\frac{1}{2}\pi^{\pm}\left(\mathbbm{1}+\mathcal{V}-4\mathcal{T}\right) \nonumber \\ \Pi^{\pm}_{\mathcal{B}}&=\frac{1}{2}\pi^{\pm}\left(\mathbbm{1}-\mathcal{V}\right)
    \end{align}
where $\pi^{\pm}=\frac{1}{2}(\mathbbm{1}\pm U)$ are projectors onto the subspaces acted on by $\mathcal{A}^{\pm}$. The angle $\zeta$ is necessary as the S-invariant Choi matrices are a non-abelian family of density matrices. The projectors onto the trivial representations are generated by the pure states,
\begin{align}
    |\Phi^{+}_{\zeta}\rangle&=\frac{\cos{\frac{\zeta}{2}}}{\sqrt{p}} |\Psi_{p}\rangle+\frac{\sin{\frac{\zeta}{2}}}{\sqrt{q}}|\Psi_{q}\rangle \nonumber \\  
    |\Phi^{-}_{\zeta}\rangle&= \frac{\sin{\frac{\zeta}{2}}}{\sqrt{p}}|\Psi_{p}\rangle-\frac{\cos{\frac{\zeta}{2}}}{\sqrt{q}}|\Psi_{q}\rangle \,. 
\end{align}
For non-zero $y$, a relative phase $e^{\pm i\xi}$ should be included between the states $|\Psi_{p}\rangle$ and $|\Psi_{q}\rangle$ in the above definitions. The states $|\Phi^{\pm}_{\zeta}\rangle$ are antipodal points on the Bloch sphere lying in the $xz$-plane. The angle $\zeta$ is defined as,
\begin{align}
    \tan\zeta=\frac{{\rm x}}{{\rm z}+2({\rm u_{\rm R}}+{\rm u_{\rm L}})} \,,
\end{align}
making $[\varrho_{\mathcal{A}},\Pi^{\pm}_{\mathbbm{1}}]=0$. An interesting detail about the families of S-invariant projectors is that they all include a pair that project into the invariant space $\mathcal{B}$. As a real representation, it is irreducible, but as a complex representation, it is reducible to two components $\mathcal{B}^{\pm}$. These components must be treated equally by an S-covariant channel, as otherwise the channel would not be hermiticity preserving. However, the (super) unitary operator that diagonalizes $\varrho_{\mathcal{A}}$ does not need to preserve the real invariant subspaces, hence the appearance of both invariant projectors $\Pi^{\pm}_{\mathcal{B}}$. Using these S-invariant projectors, the completely positive constraints are found to be 
\begin{align}
\lambda^{\pm}_{\mathbbm{1}}&={\rm w}+{\rm v}+\frac{1}{2}{\rm t}\pm\frac{1}{2}\sqrt{{\rm x}^{2}+({\rm z}+2({\rm u_{\rm R}}+{\rm u_{\rm L}}))^{2}}\geq0  \nonumber \\
\lambda^{\pm}_{\mathcal{A}}&={\rm w}\pm ({\rm u_{\rm R}}+{\rm u}_{\rm L})+{\rm v}\geq0 \nonumber \\
\lambda^{\pm}_{\mathcal{B}}&={\rm w}\pm( {\rm u_{\rm R}}-{\rm u_{\rm L}})-{\rm v} \geq0 \,,
\end{align}
where the trace-preserving conditions have not been explicitly imposed. 

Unfortunately, no simple transformation of parameters yields the completely co-positive constraints. Instead, we need to directly construct the $\widetilde{\rm S}$-invariant projectors To construct these projectors, we define the identity operators
\begin{align}
\mathbbm{1}_{\mathcal{P}}&=\sum_{\ell_{+},m_{+}}|\ell_{+}m_{+}\rangle\!\langle \ell_{+}m_{+}| \nonumber \\
\mathbbm{1}_{\mathcal{Q}}&=\sum_{\ell_{-},m_{-}}|\ell_{-}m_{-}\rangle\!\langle \ell_{-}m_{-}| \,.
\end{align}
Using the algebra of $\widetilde{\rm S}$-invariant, found in Appendix \ref{AppC}, we construct the following $\widetilde{\rm S}$-invariant projectors
\begin{align}
\Pi^{\pm}_{\mathcal{P}}&=\frac{1}{2}\left(\mathbbm{1}_{\mathcal{P}}\pm\widetilde{P}\right) \nonumber \\
\Pi^{\pm}_{\mathcal{Q}}&=\frac{1}{2}\left(\mathbbm{1}_{\mathcal{Q}}\pm\widetilde{Q}\right) \nonumber \\ 
\Pi^{\pm}_{\mathcal{R}}&=\frac{1}{4}\left(\left(\mathbbm{1}-\mathcal{V}\right)\pm\left(2\cos{\widetilde{\zeta}}(\widetilde{\mathcal{R}}+\widetilde{\mathcal{R}}^{\dag})+\sin{\widetilde{\zeta}}(\mathcal{U}_{\rm R}-\mathcal{U}_{\rm L})\right)\right) \,.
\end{align}
The pairs  $\Pi^{\pm}_\mathcal{P}$ and $\Pi^{\pm}_\mathcal{Q}$ project onto the symmetric and anti-symmetric parts of the $\mathcal{A}^{+}\otimes\mathcal{A}^{+}$ and $\mathcal{A}^{-}\otimes\mathcal{A}^{-}$ respectively. This is evident since $\widetilde{\mathcal{P}}$ and $\widetilde{\mathcal{Q}}$ are the flip operators over their respective subspaces. The parameter dependence, of the final pair of projectors, is determined by setting $[\widetilde{\varrho}_{\mathcal{A}},\Pi^{\pm}_{\mathcal{R}}]=0$, enforcing $\widetilde{\zeta}$ be defined as, 
\begin{align}
\tan{\widetilde{\zeta}}=\frac{{\rm u_{\rm R}}-{\rm u_{\rm L}}}{\frac{\rm x}{\sqrt{4pq}}} \,. 
\end{align}
These are projectors onto orthogonal mixtures of the $\mathcal{B}^{\pm}\otimes \mathcal{B}^{\pm}$ subspaces, as $\Pi_{\mathcal{B}}=\Pi^{+}_{\mathcal{R}}+\Pi^{-}_{\mathcal{R}}$ and tr$\Pi^{\pm}_{\mathcal{R}}=pq$ for any $\widetilde{\zeta}$. From this set of projectors we obtain the completely co-positive constraints, 
\begin{align}
\label{eq:SCCP}    \lambda^{\pm}_{\mathcal{P}}&={\rm w}+\left({\rm u}_{\rm R}+{\rm u}_{\rm L}\right)+{\rm v}\pm \frac{1}{2p}({\rm t}+{\rm z})\geq 0 \nonumber \\
    \lambda^{\pm}_{\mathcal{Q}}&={\rm w}-\left({\rm u}_{\rm R}+{\rm u}_{\rm L}\right)+{\rm v}\pm \frac{1}{2q}({\rm t}-{\rm z})\geq 0 \nonumber \\
    \lambda^{\pm}_{\mathcal{R}}&={\rm w}-{\rm v}\pm \sqrt{({\rm u}_{\rm R}-{\rm u}_{\rm L})^{2}+\frac{{\rm x}^{2}}{4pq}} \geq0 \,.
\end{align}
Ostensibly, there are 12 conditions for an S-covariant map to be a PPT channel. However, not all of these conditions are independent. Assuming we have an S-channel, then it follows that, 
\begin{align}
\lambda^{+}_{\mathbbm{1}}+\lambda^{-}_{\mathbbm{1}}=\left(\lambda^{+}_{\mathcal{A}}+\frac{{\rm t}+{\rm z}}{2}\right)+\left(\lambda^{-}_{\mathcal{A}}+\frac{{\rm t}-{\rm z}}{2}\right) \geq 0 \,.
\end{align}
Further, it follows from the product $\lambda^{+}_{\mathbbm{1}}\lambda^{-}_{\mathbbm{1}}\geq0$ that
\begin{align}
\label{eq:prodpos}
\left(\lambda^{+}_{\mathcal{A}}+\frac{{\rm t}+{\rm z}}{2}\right)\left(\lambda^{-}_{\mathcal{A}}+\frac{{\rm t}-{\rm z}}{2}\right)\geq \frac{{\rm x}^{2}}{4}\geq 0\, 
\end{align}
implying that $\left(\lambda^{\pm}_{\mathcal{A}}+\frac{{\rm t}\pm{\rm z}}{2}\right) \geq 0$. Noting that $\lambda^{\pm}_{\mathcal{A}}\geq0$, we can conclude that
\begin{align}
\lambda^{+}_{\mathcal{P}}&= \frac{1}{p}\left(p \lambda^{+}_{\mathcal{A}}+\frac{{\rm t}+{\rm z}}{2}\right) \geq0  \nonumber \\
\lambda^{+}_{\mathcal{Q}}&= \frac{1}{q}\left(q \lambda^{-}_{\mathcal{A}}+\frac{{\rm t}-{\rm z}}{2}\right) \geq0 
\end{align}
for any S-channel. The completely co-positive condition derived from $\lambda^{+}_{\mathcal{R}}$ is also redundant as, 
\begin{align}
\lambda^{+}_{\mathcal{R}}=\lambda^{+}_{\mathcal{B}}+\lambda^{-}_{\mathcal{B}}+\sqrt{({\rm u}_{\rm R}-{\rm u}_{\rm L})^{2}+\frac{{\rm x}^{2}}{4pq}} \geq0 
\end{align}
automatically follows from $\lambda^{\pm}_{\mathcal{B}}\geq0$. The remaining 3 completely co-positive conditions are genuinely distinct, yielding a total of 9 constraints.

In \cite{Singh_2022} it was shown that all diagonal unitary channels channels obey the PPT$^{2}$-conjecture. Therefore, since S-type unitary group covariance implies diagonal unitary covariance, we already know that S-channels also satisfy this conjecture. We are interested in showing whether the conjecture is trivially true (i.e. does EB=PPT), or showing that there exist PPT entangled S-invariant states. To do so, it is helpful to express the action of $\mathscr{E}_{\mathcal{A}}$ using the results of \cite{Singh_2021}. In this work it was shown that all diagonal unitary covariant channels take the form, 
\begin{align}
    \mathscr{E}_{\rm DUC}(\mathcal{O})={\rm diag}({\rm F}|{\rm diag} \mathcal{O}\rangle)+\underbar{{\rm G}}\odot \mathcal{O}
\end{align}
where the first term is a classical channel that mixes the diagonal entries (or populations) of the input with F being an $n\times n$ complex matrix.
The second term involves the Hadamard product ($\odot$) i.e. the entry-wise multiplication of matrices. $\underbar{{\rm G}}$ denotes the off-diagonal entries of the $n\times n$ complex matrix G, where diag(F)=diag(G). Note that diagonal unitary covariant channels are called conjugate diagonal unitary in \cite{Singh_2021}. The discrepancy appears as in that work covariance is defined with respect to the Choi matrix primarily, and not the quantum channel.

Letting $\mathbbm{J}_{i\times j}$ denote the $i\times j$ matrix with all entries equal to 1, we find that for $\mathscr{E}_{\rm DUC}=\mathscr{E}_{\mathcal{A}}$ we have, 
\begin{align}
{\rm F}&=\begin{pmatrix}
    \lambda^{+}_{\mathcal{A}} \mathbbm{J}_{p\times p}+\left(\frac{{\rm t}+{\rm z}}{2p}\right)\!\mathbbm{1}_{p} & \lambda^{+}_{\mathcal{B}}\mathbbm{J}_{p\times q} \\ \\ \lambda^{-}_{\mathcal{B}}\mathbbm{J}_{q\times p} & \lambda^{-}_{\mathcal{A}}\mathbbm{J}_{q\times q}+\left(\frac{{\rm t}-{\rm z}}{2q}\right)\!\mathbbm{1}_{q}
\end{pmatrix} \nonumber \\\nonumber\\
    {\rm G}&=\begin{pmatrix}
    \left(\frac{{\rm t}+{\rm z}}{2p}\right)\!\mathbbm{J}_{p\times p}+\lambda^{+}_{\mathcal{A}} \mathbbm{1}_{p} & \left(\frac{\rm{x}}{\sqrt{4pq}}\right)\!\mathbbm{J}_{p\times q} \\ \\ \left(\frac{\rm{x}}{\sqrt{4pq}}\right)\!\mathbbm{J}_{q\times p} &\left( \frac{{\rm t}-{\rm z}}{2q}\right)\!\mathbbm{J}_{q\times q}+\lambda^{-}_{\mathcal{A}}\mathbbm{1}_{q}
    \end{pmatrix} \,.
\end{align}
The conditions for $\mathscr{E}_{\rm DUC}$ to be completely-positive are that F be entry-wise non-negative and that G be positive semi-definite. The eigenvalues of G are precisely $\lambda^{\pm}_{\mathbbm{1}}$ and $\lambda^{\pm}_{\mathcal{A}}$, therefore these must be non-negative for G to be positive semi-definite. Once these eigenvalues are non-negative, it automatically follows that the diagonal entries of G (and hence F) are also non-negative. The remaining conditions for complete positivity are that $\lambda^{\pm}_{\mathcal{B}}\geq0$. Therefore the completely positive conditions exactly match those previously derived. Similarly, the completely co-positive constraints  F$_{ij}$F$_{ji}\geq$|{\rm G}$_{ij}|^{2}$ for all $i$ and $j$, reproduce the content of Eq.(\ref{eq:SCCP}). 

Using Lemma 6.11 from \cite{Singh_2021}, $\mathscr{E}_{\rm DUC}$ is entanglement breaking whenever (F,G) is pairwise completely-positive (PCP). That is a finite collection of vectors exist such that, 
\begin{align}
\label{eq:PCP}
    {\rm F}&=\sum_{i}|h_{i}\odot\bar{h}_{i}\rangle \langle k_{i}\odot \bar{k}_{i} |\\
    {\rm G}&= \sum_{i}|h_{i}\odot k_{i}\rangle \langle h_{i}\odot k_{i}|\,, 
\end{align}
where $i$ is some finite index. To begin, lets note that G is always diagonalizable as a symmetric matrix, therefore we have the eigendecomposition, 

\begin{align}
{\rm G}=\sum_{i}G_{i}|e_{i}\rangle \langle e_{i}|= \sum_{i}|h_{i}\odot k_{i}\rangle \langle h_{i}\odot k_{i}| \,,
\end{align}
where we assumed that the PCP decomposition, if it exists, should at least be the size of the eigendecomposition of G. Vector pairs may be added to the collection, so long as they have zero Hadamard product, without altering the value of G. The projection operators, 
\begin{align}
\Pi_{p}&=\mathbbm{1}_{p}-\frac{1}{p}\mathbbm{J}_{p} \nonumber \\ 
\Pi_{q}&=\mathbbm{1}_{q}-\frac{1}{q}\mathbbm{J}_{q} \, 
\end{align}
commute with G, as can be easily checked. Therefore, G has $(p-1)$ degenerate eigenvectors that sum to $\Pi_{q}$ and $(q-1)$ degenerate eigenvectors that sum to $\Pi_{q}$. Explicitly we have,  
\begin{align}
    {\rm G}&=\lambda^{+}_{\mathcal{A}}\Pi_{p}+\lambda^{-}_{\mathcal{A}}\Pi_{q}+{\rm g}\,,
\end{align}
where g is orthogonal to the projections $\Pi_{p}$ and $\Pi_{q}$. The matrix g has rank 2 and its column (and row) space are spanned by the pair of vectors,
\begin{align}
    |j_{p}\rangle &=\sum_{i_{+}}|i_{+}\rangle \nonumber \\ 
    |j_{q}\rangle &=\sum_{i_{-}}|i_{-}\rangle \,. 
\end{align}
To construct $\Pi_{p}$ (or $\Pi_{q}$), simply find $(p-1)$ (or $(q-1)$) orthogonal vectors that only have non-zero $i_{+}$ ($i_{-})$ indices that are orthogonal to $|j_{p}\rangle $ ($|j_{q}\rangle$). Notice now that, 
\begin{align}
    \Pi_{p}{\rm G}=\sum_{i'}G_{i'}|e_{i'}\rangle\langle e_{i'}|=\sum_{i}\Pi_{p}|h_{i}\odot k_{i}\rangle h_{i}\odot k_{i}|,
\end{align}
where in the first sum, eigenvectors orthogonal to $\Pi_{p}$ are simply removed. Looking at the final sum, we may conclude that each $|h_{i}\odot k_{i}\rangle$ lies entirely within an invariant subspace of G. If this is not the case, then off diagonal terms would appear in the RHS above that are clearly zero.

To continue, we may note that while F is not symmetric, it still commutes with $\Pi_{p}$ and $\Pi_{q}$. And it is not difficult to check that, at least in these invariant sectors, F has the same eigenvectors as G.  Explicitly we have, 

\begin{align}
    {\rm F}&=\frac{{\rm t}+{\rm z}}{2p}\Pi_{p}+\frac{{\rm t}-{\rm z}}{2q}\Pi_{q}+{\rm f} \,, 
\end{align}
and so by the same logic as before, we must have each $|h_{i}\odot \bar{h}_{i}\rangle$ and $|k_{i}\odot \bar{k}_{i}\rangle$ reside entirely within an invariant sector. This means that the PCP decomposition should be broken up into a orthogonal sum over invariant sectors. But notice by construction the vector sums in Eq.$(\ref{eq:PCP})$ produce matrices with the same diagonal elements. Therefore, when decomposing the PCP sums over invariant subspaces, the diagonal elements for each invariant sector must match. This condition is clearly stronger than the CCP conditions, as we must set, 
\begin{align}
\lambda^{+}_{\mathcal{A}}&=\frac{{\rm t}+{\rm z}}{2p}  \nonumber \\ 
\lambda^{-}_{\mathcal{A}}&=\frac{{\rm t}-{\rm z}}{2q} \,.
\end{align}
This is equivalent to saturating two of the CCP conditions i.e. equivalent to setting $\lambda^{-}_{\mathcal{P}}=\lambda^{-}_{\mathcal{Q}}=0$. For completeness, the matrix representations of f and g in the computational basis are  
\begin{align}
    {\rm f}&=\begin{pmatrix}
(\lambda^{+}_{\mathcal{A}}+\frac{{\rm t}+{\rm z}}{2p^{2}})\mathbbm{J}_{p\times p} & \lambda^{+}_{\mathcal{B}}\mathbbm{J}_{p\times q}\\ \lambda^{-}_{\mathcal{B}} \mathbbm{J}_{q\times p} &  (\lambda^{-}_{\mathcal{A}}+\frac{{\rm t}-{\rm z}}{2q^{2}})\mathbbm{J}_{q\times q}
\end{pmatrix} \nonumber \\ {\rm g}&=\begin{pmatrix}
(\frac{\lambda^{+}_{\mathcal{A}}}{p}+\frac{{\rm t}+{\rm z}}{2p})\mathbbm{J}_{p\times p} & \frac{\rm x}{\sqrt{4pq}}\mathbbm{J}_{p\times q}\\ \frac{\rm x}{\sqrt{4 pq }}\mathbbm{J}_{q\times p} &  (\frac{\lambda^{-}_{\mathcal{A}}}{q}+\frac{{\rm t}-{\rm z}}{2q})\mathbbm{J}_{q\times q} \,,
\end{pmatrix}
\end{align}
and notice that the diag(f)=diag(g)
follows from our previous conditions. At this point explicitly EB S-channels can be obtained, but we conclude here as we have shown that EB $\neq $ PPT for S-channels.

\section{Conclusions}
In this work, we have shown that all Cartan-covariant channels satisfy the PPT$^{2}$-conjecture, and we have further shown the existence of PPT entangled Cartan-covariant channels. It is not immediately clear what aspect of the compact symplectic group makes PPT entangled states exist, while such states do not exist for the special orthogonal group. On the surface compact symplectic symmetry shares many features with special orthogonal symmetry. Determining if group features are helpful in making statements about covariant channels entanglement properties certainly warrants further study. 

Another direction for further study is that of $d$-symmetric decompositions of the special unitary group. These are generalization of symmetric space decompositions, introduced in \cite{Gray1972}, where a Lie algebra automorphism exists such that $\theta^{d}=\mathbbm{1}$. This introduces a rich set of potential covariance structures to study based on $3$-symmetric (or greater) decompositions of $\mathfrak{su}(n)=\mathfrak{su}(n_{1}+n_{2}+n_{3})$. Such channels generalize the S-covariant channels considered here. An interesting aspect to study here is how various entanglement properties of channels change depending on what the branch (family of branches) that they live on. For example, it seems obvious that these properties do not depend on the order of $n_{i}$, but are there any non-trivial symmetries?

\section{Acknowledgments}
The author thanks Sarah Shandera for providing a bounty of incisive comments that greatly improved this article. 

\bibliographystyle{unsrturl}
\bibliography{2Qmaps}

\onecolumn\newpage
\appendix
\section{Symplectic channels in the Cartan basis}
\label{sec:SymPPT2}
In order to prove the PPT$^{2}$-conjecture for symplectic channels, we argue that only the product of extreme channels need to be considered. To make this more explicit, we consider the action of the symplectic channels in the Cartan basis. Since Sp$(\frac{n}{2})$ is simple and has a trivial center, the only non-trivial components of a symplectic channel are, 
\begin{align}
    \mathscr{D}_{\mathcal{A}}=\alpha \mathbbm{1}_{\mathcal{A}}\oplus \beta \mathbbm{1}_{\mathcal{B}} \,. 
\end{align}
\begin{figure}[h]
    \centering
    \includegraphics[width=0.5\linewidth]{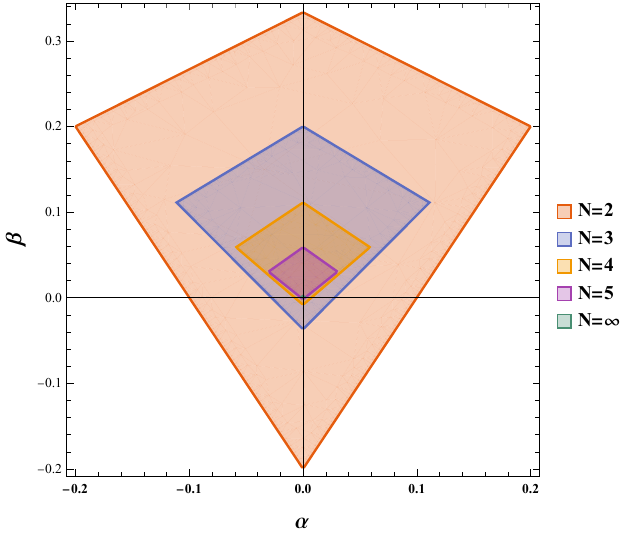}
    \caption{The region of PPT symplectic channels in the $(\alpha,\beta)$-plane are shown for several values of $n=2^{N}$. In the limit $N \rightarrow \infty$, the only PPT channel is the totally depolarizing channel, located at the origin of the $(\alpha,\beta)$-plane.}
    \label{fig:SymPPT}
\end{figure}
Computing the Choi matrix with these parameters we find, 
\begin{align}
    \varrho_{\mathcal{A}}=\frac{1-\beta}{n}\mathbbm{1}+\frac{\alpha+\beta}{2}\omega+\frac{\alpha-\beta}{2}\widetilde{\omega} \,
\end{align}
the same matrix as defined in Eq.(\ref{eq:Choisym}) with  $\Omega=\frac{\alpha+\beta}{2}$ and $\widetilde{\Omega}=\frac{\alpha-\beta}{2}$. In Fig.(\ref{fig:SymPPT}) we have plotted the regions of PPT symplectic channels in the $(\alpha,\beta)$-plane for several values of $n$. The benefit of using this picture is that parameters are multiplied under composition,
\begin{align}
  \varrho_{\mathcal{A}_{1}}\varrho_{\mathcal{A}_{2}}=\frac{1-\beta_{1}\beta_{2}}{n}\mathbbm{1}+\frac{\alpha_{1}\alpha_{2}+\beta_{1}\beta_{2}}{2}\omega+\frac{\alpha_{1}\alpha_{2}-\beta_{1}\beta_{2}}{2}\widetilde{\omega} \,.
\end{align}
The non entanglement breaking channel, with an example marked by the green dot in Fig.(\ref{fig:symPPTreg}), maps to the extreme PPT channel with $\alpha=0$ and $\beta>0$. The triangular region, generated as the convex hull of the remaining extreme channels, is the entanglement breaking region. It is obvious that if the square of the non entanglement breaking extreme channel is within the entanglement breaking region, then any other potentially interesting composition is necessarily entanglement breaking. As we proved this was the case in the article, it follows that all symplectic covariant channels obey the PPT$^{2}$-conjecture. 
\section{3-parameter SO(4)-covariant channels}

In this Appendix we prove that for a family of 3-parameter generalized SO(4)-covariant channels PPT=EB. We do this by showing such channels are equivalent to the 3-parameter channels studied in \cite{Lami_2016}. To that end, we begin with a parametrization of generalized SO(4)-covariant channels in the form given in Sec.(\ref{sec:CCQC}). To construct a Cartan Basis, we use a basis of 2-letter Pauli words $\sigma_{i_{\rm L}}\otimes \sigma_{i_{\rm R}}$. The orthogonal decomposition reads, 
\begin{align}
    \mathfrak{su}(4)=\mathcal{A}_{\rm L}\oplus \mathcal{A}_{\rm R}\oplus \mathcal{B}, 
\end{align}
with $\mathcal{A}_{\rm L}={\rm span}\left( X_{\rm L }\otimes \mathbbm{1}_{\rm R},Y_{\rm L} \otimes \mathbbm{1}_{\rm R}, Z_{\rm L }\otimes \mathbbm{1}_{\rm R}\right)$,  $\mathcal{A}_{\rm R}={\rm span}\left( \mathbbm{1}_{\rm L}\otimes X_{\rm R }, \mathbbm{1}_{\rm L}\otimes Y_{\rm R }, \mathbbm{1}_{\rm L}\otimes Z_{\rm R}\right)$, and $\mathcal{B}$ spanned by the remaining set of (genuine) 2-letter Pauli words. The center of SO(4) is trivial, therefore no non-unital parameters or unitary parameters are present. Thus, there are only 3-parameters $(\alpha_{\rm L},\alpha_{\rm R},\beta)$ which sets the amount of isotropic depolarization over each SO(4)-invariant subspace. To simplify the computation of the Choi matrix, notice that if $\mathcal{G}_{\rm i}$ is an orthonormal basis of the Killing-Cartan metric, then, 
\begin{align}
    \omega=\frac{1}{4}\sum_{\rm i}\mathcal{G}_{\rm i }\otimes \mathcal{G}^{\rm T}_{\rm i} \,, 
\end{align}
where tr$\omega=4$. 
To prove this, take a generic sum of product operators and impose the strong symmetry of the maximally entangled state. It is useful to also consider the maximally entangled states,
\begin{align}
    \omega_{\rm L_{1}L_{2}}=\frac{1}{2}\left(\mathbbm{1}_{\rm L_{1}L_{2}}+X_{\rm L_{1}}X_{\rm L_{2}}-Y_{\rm L_{1}}Y_{\rm L_{2}}+Z_{\rm L_{1}}Z_{\rm L_{2}}\right) \, 
\end{align}
and with a similar definition $\omega_{\rm R_{1}R_{2}}$. Using these definitions, and the action of the channel, we find the Choi matrix to be, 
\begin{align}
    \varrho_{\mathcal{A}}=&\frac{1}{4}\bigg(\mathbbm{1}_{\rm L_{1}L_{2}}\otimes\mathbbm{1}_{\rm R_{1}R_{2}} + +\alpha_{\rm L}\left(2\omega_{\rm L_{1}L_{2}}-\mathbbm{1}_{\rm L_{1}L_{2}}\right)\otimes\mathbbm{1}_{\rm R_{1}R_{2}}   +\alpha_{\rm R}\left(2\omega_{\rm R_{1}R_{2}}-\mathbbm{1}_{\rm R_{1}R_{2}}\right)\otimes\mathbbm{1}_{\rm L_{1}L_{2}}\nonumber  \\ &+\beta \left(2\omega_{\rm L_{1}L_{2}}-\mathbbm{1}_{\rm L_{1}L_{2}}\right)\otimes \left(2\omega_{\rm R_{1} R_{2}}-\mathbbm{1}_{\rm R_{1} R_{2}}\right) \bigg)
\end{align}
and describes a trace-preserving process since tr$\omega$=tr$\varrho_{\mathcal{A}}=4$. The Choi matrix and its partial transpose are simple enough to be diagonalized using mathematical software, such as Mathematica. Doing so we find that the region of PPT channels has 5 vertices. Four of the vertices are product channels with $(\alpha_{\rm L},\alpha_{\rm R},\beta)$-coordinates, 
\begin{align}
    \mathscr{E}_{1}&=(\frac{1}{3},\frac{1}{3},\frac{1}{9}) \nonumber \\
     \mathscr{E}_{2}&=(\frac{1}{3},-\frac{1}{3},-\frac{1}{9}) \nonumber \\
      \mathscr{E}_{3}&=(-\frac{1}{3},\frac{1}{3},-\frac{1}{9}) \nonumber \\
       \mathscr{E}_{4}&=(-\frac{1}{3},-\frac{1}{3},\frac{1}{9}) \nonumber 
       \,,
\end{align}
and the final vertex being $ \mathscr{E}_{5}=(0,0,\frac{1}{3})$, clearly not a product channel. 
 The product channels are products of depolarizing channels  $\Phi_{\pm p }\otimes \Phi_{\pm p}$ with $p=\frac{1}{3}$.  In either case we have $p\leq\frac{1}{3}$, therefore both are entanglement breaking \cite{Morav_kov__2010}. It further follows that all four extreme product PPT channels are entanglement breaking. The final extreme point is not a product channel. To show that it is entanglement breaking we use the results of \cite{Lami_2016}, where the authors introduced a three parameter generalization of bipartite depolarizing channels. They found a Choi matrix of the form, 
\begin{align}
    \Sigma_{\mathcal{A}}=\frac{1}{4}\mathbbm{1}_{\rm L_{1}L_{2}}\otimes \mathbbm{1}_{\rm R_{1}R_{2}}+\frac{\rm a_{\rm L}}{4}\mathbbm{1}_{\rm L_{1}L_{2}}\otimes\omega_{\rm R_{1}R_{2}}+\frac{\rm a_{\rm R}}{4}\omega_{\rm L_{1}L_{2}}\otimes \mathbbm{1}_{\rm R_{1}R_{2}}+\frac{\rm b}{4}\omega_{\rm L_{1} L_{2}}\otimes \omega_{\rm R_{1}R_{2}}
\end{align}
which is not trace-preserving as $\sigma$=tr$\Sigma_{\mathcal{A}}=4+2{\rm a}_{\rm L}+2{\rm a}_{\rm R}+{\rm b}$. However, if $\Sigma_{\mathcal{A}}\geq0$ then its trace is also positive. Therefore, it may be rescaled to be trace-preserving while maintaining complete positivity. Clearly both Choi matrices are linear combinations of the same observables. Therefore, we should be able to find functions $\alpha_{\rm R}=\alpha_{\rm R}({\rm a}_{\rm L},{\rm a}_{\rm R},{\rm b})$ etc that cause the Choi matrices to coalesce. To match our parametrizations we must solve the equation 
\begin{align}
    \frac{1}{\sigma}\Sigma_{\mathcal{A}}=\frac{1}{4}\varrho_{\mathcal{A}} \,. 
\end{align}
In this way we find that our extreme PPT channels map exactly to the ones found using $\Sigma_{\mathcal{A}}$. Therefore, we may conclude that EB=PPT over the set of SO(4)-covariant channels. 
\section{Algebra of 
S-type unitary invariants and dual invariants}
\label{AppC}
In this appendix we fully describe the algebra of S-invariants and $\widetilde{\rm S}$-invariants. A full set of S-invariants consists of the eight observables described in Sec.(\ref{Section S}). Four of these observables form a Pauli algebra, 
\begin{align}
   \mathcal{X}\mathcal{Y}&=-\mathcal{Y}\mathcal{X}=\frac{i}{2}\mathcal{Z} \quad \quad \mathcal{T}\mathcal{Z}=\mathcal{Z}\mathcal{T}=\frac{1}{2}\mathcal{Z} \nonumber \\ 
 \mathcal{Y}\mathcal{Z}&=-\mathcal{Z}\mathcal{Y}=\frac{i}{2}\mathcal{X} \quad \quad \mathcal{T}\mathcal{X}=\mathcal{X}\mathcal{T}=\frac{1}{2}\mathcal{X}\nonumber \\   \mathcal{Z}\mathcal{X}&=-\mathcal{X}\mathcal{Z}=\frac{i}{2}\mathcal{Y} \quad \quad \mathcal{T}\mathcal{Y}=\mathcal{Y}\mathcal{T}=\frac{1}{2}\mathcal{Y}\,,
\end{align}
with the remaining products $\mathcal{X}^{2}=\mathcal{Y}^{2}=\mathcal{Z}^{2}=\mathcal{T}^{2}=\frac{1}{2}\mathcal{T}$. The algebra of S-invariants, not including $\mathcal{X}$ or $\mathcal{Y}$, is abelian. So, to fully describe the algebra, we need the remaining products involving $\mathcal{X}$ and $\mathcal{Y}$, 
\begin{align}
  \mathcal{U}_{\rm R}\mathcal{X}&=\mathcal{U}_{\rm L}\mathcal{X}=-\mathcal{X}\mathcal{U}_{\rm R}=-\mathcal{X}\mathcal{U}_{\rm L}=i\mathcal{Y} \quad \mathcal{X}\mathcal{V} =\mathcal{V}\mathcal{X}=\mathcal{X}  \nonumber \\ 
  \mathcal{U}_{\rm R}\mathcal{Y}&=\mathcal{U}_{\rm L}\mathcal{Y}=-\mathcal{Y}\mathcal{U}_{\rm R}=-\mathcal{Y}\mathcal{U}_{\rm L}=-i\mathcal{X} \quad \mathcal{Y}\mathcal{V} =\mathcal{V}\mathcal{Y}=\mathcal{Y} \,
\end{align}
and for completeness the squares $\mathcal{U}^{2}_{\rm R}=\mathcal{U}^{2}_{\rm L}=\mathcal{V}^{2}=\mathbbm{1}$. As an example, we can show that $\Pi_{\mathcal{A}}$ (as defined in Sec.(\ref{Section S}) is a projection operator. We have, 
\begin{align}
\Pi^{2}_{\mathcal{A}}=\frac{1}{4}(\mathbbm{1}+\mathcal{V}-4\mathcal{T})(\mathbbm{1}+\mathcal{V}-4\mathcal{T})=\frac{1}{4}\left(2\mathbbm{1}+2\mathcal{V}-8\mathcal{T}\right)=\Pi_{\mathcal{A}}\,.
\end{align}
 It should be clear why we have decided to parametrize the S-channels this way, as opposed to using the parametrization found in Sec.(\ref{sec:CCQC}). The algebra of S-invariants we have constructed is much simpler that the natural algebra encountered in the direct channel parametrization (Casimir invariants). One benefit of parameterizing the S-channels directly is that the trace-preserving and unitality conditions are trivial. Letting $\mathscr{U}_{\rm R}$ and $\mathscr{U}_{\rm L}$ be the trace-less parts of $\mathcal{U}_{\rm R}$ and $\mathcal{U}_{\rm L}$, then an S-covariant channel is trace-preserving iff the coefficient in front of $\mathscr{U}_{\rm L}$ is zero. Similarly, an S-covariant channel is only unital if the coefficient in front of $\mathscr{U}_{\rm R}$ is equal to zero. 

 For the $\widetilde{\rm S}$-invariants, we only consider the products necessary to show that $\Pi^{\pm}_{\mathcal{R}}$ are projection operators. It is a straight forward matter to see that the other defined $\widetilde{\rm S}$-invariant operators are projectors and commute with the Choi matrix. In fact, we will only need the following relations, 
\begin{align}
&\widetilde{\mathcal{R}}^{\dag}\widetilde{\mathcal{R}}+\widetilde{\mathcal{R}}\widetilde{\mathcal{R}}^{\dag}=\frac{1}{2}\left(\mathbbm{1}-\mathcal{V}\right) \nonumber \\
&\widetilde{\mathcal{R}}\mathcal{U}_{\rm R}=\widetilde{\mathcal{R}}\mathcal{U}_{\rm L}=-\mathcal{U}_{\rm R}\widetilde{\mathcal{R}}=-\mathcal{U}_{\rm L}\widetilde{\mathcal{R}}=\mathcal{V}\widetilde{\mathcal{R}}=\widetilde{\mathcal{R}}\mathcal{V}=-\widetilde{\mathcal{R}}\,.
 \end{align} 
From these relations it follows that $(\Pi^{\pm}_{\mathcal{R}})^{2}=\Pi^{\pm}_{\mathcal{R}}$.  
\end{document}